\documentclass[american,aps,pra,reprint, superscriptaddress]{revtex4-1}
\usepackage{tikz}
\usepackage{amsmath,amscd,amsthm,amssymb}
\usepackage{mathrsfs}
\usepackage{graphicx}
\usepackage{epsf,epstopdf}
\usepackage[]{caption3}
\usepackage[all]{xy}
\usepackage[unicode=true,pdfusetitle, bookmarks=true,bookmarksnumbered=false,bookmarksopen=false, breaklinks=false,pdfborder={0 0 0},backref=false,colorlinks=false] {hyperref}
\hypersetup{ colorlinks,linkcolor=myurlcolor,citecolor=myurlcolor,urlcolor=myurlcolor}
\usepackage{graphics,epstopdf,graphicx, amsthm, amsmath, amssymb, times, braket, color, bm}
\usepackage[up]{subfigure}
\usepackage{cleveref}

\definecolor{myurlcolor}{rgb}{0,0,0.7}

\theoremstyle{plain}
\newtheorem{thm}{\protect\theoremname}

\providecommand{\theoremname}{Theorem}
\newcommand*{\myproofname}{Proof}

\makeatother

\newtheorem*{cor}{Corollary}

\theoremstyle{definition}
\newtheorem{defn}{Definition}%[section]

\theoremstyle{remark}
\newtheorem{rem}{Remark}

\begin{document}

 \author{Chunhe Xiong}
 \email{xiongch@zucc.edu.cn}
 \affiliation{Institute of Quantum Computing and Computer Theory, School of Computer Science and Engineering, Sun Yat-sen University, Guangzhou 510006, China}

\affiliation{Interdisciplinary Center for Quantum Information, Department of Physics, Zhejiang University, Hangzhou 310027, China}
\affiliation{School of Computer and Computing Science, Zhejiang University City College, Hangzhou 310015, China}

 \author{Sunho Kim}
 \email{kimsunho81@hrbeu.edu.cn}
 \affiliation{School of Mathematical Sciences, Harbin Engineering University, Harbin 150001, China}

 \author{Daowen Qiu}
 \email{issqdw@mail.sysu.edu.cn}
\affiliation{Institute of Quantum Computing and Computer Theory, School of Computer Science and Engineering, Sun Yat-sen University, Guangzhou 510006, China}
\affiliation{Instituto de Telecomunica\c{c}{\H o}es, Departamento de Matem{\' a}tica, Instituto Superior T{\' e}cnico, Av. Rovisco Pais 1049-001, Lisbon, Portugal}

 \author{Asutosh Kumar}
 \email{asutoshk.phys@gmail.com}
 \affiliation{P.G. Department of Physics, Gaya College, Magadh University, Rampur, Gaya 823001, India}

 \affiliation{Harish-Chandra Research Institute, HBNI, Chhatnag Road, Jhunsi, Allahabad 211019, India}
\affiliation{Vaidic and Modern Physics Research Centre, Bhagal Bhim, Bhinmal, Jalore 343029, India}

 \author{Junde Wu}
\email{wjd@zju.edu.cn}
 \affiliation{School of Mathematical Sciences, Zhejiang University, Hangzhou 310027, PR~China}

\title{Characterizing entanglement using quantum discord over state extensions}
%\title{Entanglement as the minimal quantum discord over state extensions}

\begin{abstract}
We propose a framework to characterize entanglement with quantum discord, both asymmetric and symmetric, over state extensions. In particular, we show that the minimal Bures distance of discord over state extensions is equivalent to Bures distance of entanglement. This equivalence places quantum discord at a more primitive position than entanglement conceptually in the sense that entanglement can be interpreted as an irreducible part of discord over all state extensions. Based on this equivalence, we also offer an operational meaning of Bures distance of entanglement by connecting it to quantum state discriminations. Moreover, for the relative entropy part, we prove that the entanglement measure introduced by Devi and and Rajagopal [A. R. U. Devi and A. K. Rajagopal, Phys. Rev. Lett. {\bf 100}, 140502 (2008)] is actually equivalent to the relative entropy of entanglement. We also provide several quantifications of entanglement based on discord measures.

\end{abstract}
\maketitle

\section{introduction}

Quantum correlations \cite{horodeckiRMP09,kavanRMP2012} are defined from different viewpoints and they, in turn, are expected to offer different advantages. Hence, the characterization and quantification of quantum correlations is instrumental in exploring and exploiting the quantum phenomena. The remarkable advantages that quantum correlations offer make quantum information theory more powerful than classical theory.
Entanglement \cite{horodeckiRMP09} is an important quantum resource which plays a crucial role in quantum information processing, quantum algorithms, quantum computation and cryptography \cite{nielsen10}. The notion of quantum correlations and resources beyond entanglement \cite{kavanRMP2012,bartlettRMP2007,chitambarRMP2019,streltsovRMP2017} such as quantum discord \cite{kavanRMP2012,zurek2001a,vedral2001a,rulli2011,xu2013}, quantum coherence \cite{streltsovRMP2017, baumgratz2014, streltsov2017, saxena2020}, etc. are also very prominent and useful in quantum information theory. For instance, quantum discord is the genuine resource in the DQC1 algorithm \cite{datta2005,datta2008a}.

Among several quantum correlations, entanglement and quantum discord are significant, and are usually regarded as very distinct in nature. While entanglement belongs to the entanglement-separability paradigm, quantum discord belongs to the information-theoretic paradigm.
Nevertheless, entanglement and quantum discord have both essential similarities and significant differences. Several remarkable investigations 
\cite{cubbit2003,koashi2004,Cornelio2011,Fanchini2011,cenlx2011,adesso2010,cavalcanti2011,piani2011,
madhok2011,streltsov2011,streltsov2012,piani2012,chuan2012}
relating to entanglement and discord have been studied in recent years.
Especially, the correspondence between classical states versus separable states \cite{linan2008}, the characterization and quantification of entanglement with the generalized information-theoretic measure \cite{devi2008}, the minimal quantum discord of bipartite state over state extensions \cite{luo2016} and the minimal correlated coherence over symmetric state extensions \cite{tan2016,tan2018} are of special interest.

It should be noted that the framework that quantifies entanglement with quantum discord or coherence over state extensions is quite different from the existing entanglement measures such as entanglement of formation and entanglement cost \cite{bennett1996}, distillable entanglement \cite{bennett1996c}, relative entropy of entanglement and Bures distance of entanglement \cite{vedral1997,vedral1998}, robustness of entanglement \cite{vidal1999a}, and squashed entanglement \cite{tucci2002,christandl2004},
which are mostly based on operational meaning, information principles, and mathematics.  Thus far, the interrelationship between these two different categories of entanglement measures is not clear.

In this paper, we study {\it characterization of entanglement using quantum discord over state extensions}. 
For what do we mean by ``over state extensions'', see \cite{note-state-extension}. 
Our work is different from and generalisation of previous works 
\cite{devi2008, luo2016, tan2016}.
Authors in Ref. \cite{devi2008} introduced {\it quantumness of correlations} as an entanglement measure. They proved that for a bipartite state $\rho_{ab}$, the relative entropy of asymmetric discord over all extended states is a candidate of entanglement measure, that is, this quantification is faithful for separable states and non-increasing under local operations. It also turns out to be an upper bound to the relative entropy of entanglement.
On the other hand, it was shown in Ref. \cite{luo2016} that the minimal discord over all extended states is also a valid entanglement measure. These two investigations revealed the potential relation between entanglement and quantum discord over state extensions. 
Also, the conservation law for distributed entanglement and quantum discord \cite{Fanchini2011} hints at some kind of fundamental relation between them. 
However, a general formalism relating entanglement and discord (quantum correlations regarded as very distinctive in nature) has been elusive.
Our study provides a natural connection between the two via state extensions. The main findings of our work can be summarised as follows.\\ 
%The Bures metric \cite{Bures1969,Uhlmann1976} provides a nice distance on the convex cone of density matrices. In particular, Bures distance is monotonous, Riemannian \cite{Petz1996} and its metric coincides with the quantum Fisher information which plays an important role in high precision interferometry \cite{Braunstein1994}.
(i) For Bures distance \cite{bures1969,uhlmann1976}, we prove that the minimal Bures distance of discord over state extensions is equivalent to Bures distance of entanglement \cite{vedral1998,wei2003,streltsov2010}. 
Moreover, Bures distance of entanglement was proved to be equal to its corresponding convex roof \cite{streltsov2010}, which plays a key role in our study. In fact, we prove the equivalence by showing that the minimal Bures distance of discord over state extensions is bounded by the Bures distance of entanglement and its convex roof.

Bures distance of entanglement is defined from the geometric viewpoint whose operational meaning is not very clear.  A correspondence, however, between Bures distance of discord and  quantum state discriminations has been established in Ref.  \cite{spehner2013a}. Based on this correspondence and the equivalence between Bures distance of entanglement and discord over state extensions, we offer an operational meaning of Bures distance of entanglement by linking it to quantum state discriminations.

(ii) We propose a framework to characterize entanglement using quantum discord over state extensions. Actually, for a generalized discord measure including entropic discord \cite{zurek2001a, vedral2001a}, geometric discord \cite{dakic2010}, measurement-induced geometric discord \cite{luo2008b}, correlated coherence \cite{tan2016} and geometric correlated partial coherence \cite{xiong2019b}, we prove that the minimal discord over state extensions is a candidate of an entanglement measure. This provides an alternative perspective to understand entanglement from the viewpoint of quantum discord. Results in previous studies \cite{devi2008, luo2016, tan2016} are special cases of our framework.

(iii) We show that quantification of entanglement proposed in Ref. \cite{devi2008} is actually equivalent to the relative entropy of entanglement \cite{vedral1998}, which is a well-studied entanglement measure.
%
%For relative entropy, we prove that the corresponding discord measure is equal to the measurement-induced discord, which implies that the quantifier of entanglement proposed in Ref.  \cite{devi2008} is equivalent to the relative entropy of entanglement \cite{vedral1998}.

This paper is structured as follows. In Sec. \ref{prelims}, we recall various concepts prerequisite for our study. We discuss characterization of entanglement using asymmetric and symmetric quantum discord in Sec. \ref{characterization-asymmetric} and Sec. \ref{characterization-symmetric} respectively. We give an operational meaning of the Bures distance of entanglement in Sec. \ref{characterization-asymmetric} C. We conclude our findings in Sec. \ref{conclusion}. Proofs of some theorems and symbols with their meanings are presented in Sec. \ref{appendix} (Appendix) at the end of the bibliography.

\section{Preliminaries}\label{prelims}
In this section, we briefly recall various concepts which are prerequisite for our study.

\subsection{Separable states}
Let $\mathcal{H}=\mathcal{H}_a\otimes\mathcal{H}_b$ be the composite Hilbert space of a bipartite system, and $D(\mathcal{H})$ be the
set of density matrices on $\mathcal{H}$. A quantum state $\rho_{ab}\in D(\mathcal{H})$ shared between two parties \(a\) and \(b\) is called separable if it can be represented as
\begin{align}
\rho_{ab}=\sum_ip_i\rho^i_a\otimes\rho^i_b,
\end{align}
where $p_i\ge0, \sum_ip_i=1$ and $\rho^i_a$, $\rho^i_b$ are local states for parties \(a\) and \(b\). Otherwise, it is called entangled. We denote the set of separable states by $\mathcal{S}$.
%The separable states are prepared by local (quantum) operations and classical communications (LOCC). %That is, to prepare a separable state, Alice first prepares a state $\rho^i_a$ with probability $p_i$ and communicates this to Bob. Based on this information, Bob prepares the corresponding state $\rho^i_b$. However, it is not possible to create entangled states in this way.
Moreover, a bipartite separable state $\rho_{ab}$ is called classical-quantum ($\mathcal{CQ}$) if it can be written as $\rho_{ab}=\sum_ip_i\ket{i}_a\bra{i}\otimes\rho^i_b$, and classical-classical ($\mathcal{CC}$) if $\rho_{ab}=\sum_{i,j}p_{ij}\ket{i}_a\bra{i}\otimes\ket{j}_b\bra{j}$, where $\{\ket{i}\}, \{\ket{j}\}$ are two set of orthogonal pure states.
%%\begin{eqnarray*}
%$\rho_{ab} = \sum_{i,j}(\Pi^i_a\otimes\Pi^j_b)\rho_{ab}(\Pi^i_a\otimes\Pi^j_b)
%= \sum_{i,j}p_{ij}\ket{i}_a\bra{i}\otimes\ket{j}_b\bra{j}$,
%%\end{eqnarray*},where $\{\ket{i}\}$ is a set of orthogonal pure states.
%\begin{eqnarray*}

%\end{eqnarray*}
% classified into following types:
%(i) classical-classical ($\mathcal{CC}$) if
%%\begin{eqnarray*}
%$\rho_{ab} = \sum_{i,j}(\Pi^i_a\otimes\Pi^j_b)\rho_{ab}(\Pi^i_a\otimes\Pi^j_b)
%= \sum_{i,j}p_{ij}\ket{i}_a\bra{i}\otimes\ket{j}_b\bra{j}$,
%%\end{eqnarray*}
%(ii) classical-quantum ($\mathcal{CQ}$) if
%%\begin{eqnarray*}
%$\rho_{ab} = \sum_{i}(\Pi^i_a\otimes I_b)\rho_{ab}(\Pi^i_a\otimes I_b)
%= \sum_ip_i\ket{i}_a\bra{i}\otimes\rho^i_b$,
%%\end{eqnarray*}
%and (iii) quantum-classical ($\mathcal{QC}$) if
%%\begin{eqnarray*}
%$\rho_{ab} = \sum_{j}(I_a \otimes \Pi^j_b)\rho_{ab}(I_a \otimes \Pi^j_b)
%= \sum_j p_j \rho^j_a \otimes \ket{j}_b\bra{j}$,
%%\end{eqnarray*}
%where the probabilities are nonnegative and sum to unity, $\{\Pi^i_a\}$ and $\{\Pi^j_b\}$ are von Neumann measurements, $\{\ket{i}_a\}$ and $\{\ket{j}_b\}$ are orthonormal basis of parties \(a\) and \(b\), respectively.
%%Otherwise, a separable state is called quantum-quantum ($\mathcal{QQ}$).

\subsection{Relative entropy}

Relative entropy is defined as $S(\rho||\sigma):=\textmd{tr}[\rho(\log\rho-\log\sigma)]$ \cite{nielsen10}.
Throughout, unless stated otherwise, the base of the $\log$ should be taken \(2\).
For any classical-quantum state $\sigma_{ab}^{(cq)}=\sum_jp_j\ket{j}_a\bra{j}\otimes\sigma^j_{b}$, it is easy to verity that $\textmd{tr}(\rho_{ab}\log\sigma_{ab}^{(cq)})=\textmd{tr}[\Pi_a(\rho_{ab})\log\sigma_{ab}^{(cq)}],$
where $\Pi_a(\rho_{ab})=\sum_i(\ket{i}_a\bra{i}\otimes I_b)\rho_{ab}(\ket{i}_a\bra{i}\otimes I_b)$. As a result, one has
\begin{align}
&S(\rho_{ab}||\sigma_{ab}^{(cq)})=\textmd{tr}[\rho_{ab}\log\rho_{ab}]-\textmd{tr}[\rho_{ab}\log\sigma_{ab}^{(cq)}] \nonumber \\
%=&\textmd{tr}[\rho_{ab}\log\rho_{ab}]-\textmd{tr}[\Pi_a(\rho_{ab})\log\Pi_a(\rho_{ab})] \nonumber \\
%+&\textmd{tr}[\Pi_a(\rho_{ab})\log\Pi_a(\rho_{ab})]-\textmd{tr}[\Pi_a(\rho_{ab})\log\sigma_{ab}^{(cq)}] \nonumber \\
=&S(\Pi_a(\rho_{ab}))-S(\rho_{ab})+S(\Pi_a(\rho_{ab})||\sigma_{ab}^{(cq)}).
\label{eq-relent}
\end{align}
We remark that a similar equation is established in block coherence theory \cite{aberg2006}.

%Actually, if we assume $\sigma_{ab}^{(cq)}=\sum_{jk}q_{jk}\ket{j}_a\bra{j}\otimes\ket{k;j}_a\bra{k;j}$, then
%\begin{align}
%&\textmd{tr}[\Pi_a(\rho_{ab})\log\sigma_{ab}^{(cq)}] \nonumber \\
%=&\sum_{jkl}\log{q_{jk}}\textmd{tr}[(\ket{l}_a\bra{l}\otimes I_b)\rho_{ab}(\ket{l}_a\bra{l}\otimes I_b)\ket{j}_a\bra{j}\otimes\ket{k;j}_a\bra{k;j}] \nonumber \\
%=&\textmd{tr}[\rho_{ab}\sum_{jk}\log{q_{jk}}\ket{j}_a\bra{j}\otimes\ket{k;j}_a\bra{k;j}]=\textmd{tr}[\rho_{ab}\log\sigma_{ab}^{(cq)}].
%\end{align}

%Hence,

\subsection{Bures distance}
The Bures distance is defined as \cite{bures1969,uhlmann1976}
\begin{align}\label{Bures-distance}
d_B(\rho,\sigma):=\sqrt{2-2F(\rho,\sigma)},
\end{align}
where $F(\rho,\sigma)$ is the fidelity $F(\rho,\sigma):=tr\sqrt{\sqrt{\sigma}\rho\sqrt{\sigma}}$ between $\rho$ and $\sigma$. Since $F(\rho,\sigma)\in[0,1]$ and is unity iff $\rho=\sigma$, $d_B(\rho,\sigma)$ is nonnegative and vanishes iff $\rho=\sigma$. Moreover, the monotonicity and joint concavity of fidelity \cite{nielsen10} implies that $d^2_B$ is contractive and jointly convex.

Unlike relative entropy, Bures distance is a bona fide distance on state space, that is, it is faithful, symmetric, and satisfies the triangle inequality. Moreover, it is Riemannian \cite{spehner2014} and its metric is given in \cite{Hubner1992,Sommers2003,spehner2014}. 
Quantum Fisher information--a quantum analog of Fisher information in statistics--is defined in terms of the Bures distance metric \cite{Braunstein1994}. These properties of Bures distance enable it to quantify various quantum correlations including entanglement \cite{vedral1997}, discord \cite{spehner2013a,aaronson2013}, coherence \cite{baumgratz2014}, etc.

\subsection{Entanglement}
An entanglement measure $E$ is a functional on $D(\mathcal{H})$ satisfying (1) Faithfulness: $E(\rho)\ge0$, where the equality holds iff $\rho\in\mathcal{S}$, and (2) Monotonicity: $E(\rho)\ge E(\Phi(\rho))$ for any LOCC operation $\Phi$.\\

{\it Bures distance of entanglement}.--The Bures distance of entanglement is defined as the minimal square of Bures distance to separable states \cite{vedral1997},
\begin{align}\label{B-discord}
E_B(\rho_{ab}):=\min_{\sigma_{ab}\in\mathcal{S}}d^2_B(\rho_{ab},\sigma_{ab}).
\end{align}
Obviously, $E_B(\rho_{ab})$ is nonnegative, and vanishes iff $\rho_{ab}$ is separable. Furthermore, $E_B$ is convex and non-increasing under LOCC operations \cite{bennett1996, vedral1998}.
Note that for the bipartite pure state $\ket{\psi}=\sum_i\sqrt{\lambda_i}\ket{x_i}_a\ket{y_i}_b$ with $\lambda_1\ge...\ge\lambda_n$, $E_B(\ket{\psi})=2(1-\sqrt{\lambda_1})$ \cite{streltsov2010}. In fact, if we assume that $\sigma_{ab}\in\mathcal{S}$ has a separable pure state decomposition $\sigma_{ab}=\sum_jq_j\ket{\phi_j}_{ab}\bra{\phi_j}$, then
\begin{align}
&F(\ket{\psi},\sigma_{ab})=\sqrt{\sum_jq_j |\bra{\psi}\phi_j\rangle|^2}\nonumber\\
\le&\sqrt{\sum_jq_j|\bra{\psi}\phi_{max}\rangle|^2}=|\bra{\psi}\phi_{max}\rangle|,
\end{align}
where $|\bra{\psi}\phi_{max}\rangle| := \max_{j} |\bra{\psi}\phi_j\rangle|$ is the maximal over all $j$. Therefore, $E_B(\ket{\psi})=2(1-\sqrt{\lambda_1})$ and the corresponding closest separable state is $\ket{x_1,y_1}_{ab}$.\\

{\it Convex roof of Bures distance of entanglement}.--The convex roof of Bures distance of entanglement is defined as
\begin{align}
E^{cr}_B(\rho_{ab}):=\min_{p_i,\ket{\psi_i}}\sum_ip_iE_B(\ket{\psi_i}),
\end{align}
where the minimal is taken over all pure state decompositions $\rho_{ab}=\sum_ip_i\ket{\psi_i}\bra{\psi_i}$.

\subsection{Quantum discord}
{\it Asymmetric quantum discord}.--A functional $\hat{D}$ on $D(\mathcal{H})$ is called a discord measure (asymmetric or local) if it satisfies the following properties:

(D1) $\hat{D}$ is faithful, i.e., $\hat{D}(\rho_{ab})\ge0$ and the equality holds if and only if $\rho_{ab}\in\mathcal{CQ}$.

(D2) $\hat{D}$ is non-increasing for any quantum operation on subsystem \(b\), i.e., $\hat{D}(\rho_{ab})\ge \hat{D}(I_a \otimes \Phi_b(\rho_{ab}))$ for any local operation $\Phi_b$.

(D3) $\hat{D}$ is invariant under local unitary transformations, i.e., $\hat{D}(U_a\otimes U_b \rho_{ab} U^{\dagger}_a\otimes U^{\dagger}_b)=\hat{D}(\rho_{ab})$ for any unitary $U_a\otimes U_b$ acting on $\mathcal{H}_a \otimes \mathcal{H}_b$.

(D4) $\hat{D}$ reduces to an entanglement monotone \cite{vidal2000} for pure states.

Quantum discord has been studied extensively from different viewpoints in the forms of geometric discord, measurement-induced geometric discord \cite{roga2016} and geometric correlated partial coherence \cite{xiong2019b}. Quantum discord, obviously, captures quantum correlation beyond entanglement in the sense that a separable state may have nonzero quantum discord.\\

{\it Symmetric quantum discord}.--A functional $\tilde{D}$ in state space is called (symmetric or global) discord \cite{rulli2011,xu2013} measure if it satisfies the following properties:

(D${'}$1) $\tilde{D}$ is faithful, i.e., $\tilde{D}(\rho_{ab})\ge0$ and the equality holds if and only if $\rho_{ab} \in\mathcal{CC}$.

%(2) $D_a$ is non-increasing in quantum operation in subsystem b, i.e., $D^a(\rho)\ge D^a(I\otimes\Phi_b(\rho))$ for any local operation $\Phi_b$.

(D${'}$2) $\tilde{D}$ is invariant under local unitary transformations, i.e., $\tilde{D} (U_a\otimes U_b \rho_{ab} U^{\dagger}_a\otimes U^{\dagger}_b)=\tilde{D}(\rho_{ab})$ for any unitary acting on $\mathcal{H}_a$ and $\mathcal{H}_b$.

(D${'}$3) $\tilde{D}$ reduces to an entanglement monotone on pure states.\\

{\it Bures distance of discord}.--The Bures distance of discord is defined as the minimal square of Bures distance to classical-quantum states \cite{aaronson2013,spehner2013a},
\begin{align}
\hat{D}_B(\rho_{ab}):=\min_{\sigma_{ab}\in\mathcal{CQ}}d^2_B(\rho_{ab},\sigma_{ab}).
\end{align}
It has been shown that $\hat{D}_B$ is a discord measure, i.e., it satisfies the properties (D1-D4).

\begin{rem}
%It is worthy noting that that Quantum Fisher information has been introduced by Braunstein and Caves, is  related it
%to the metric - called the ``distinguishability metric" by Wootters \cite{Wootters1981}. Recently, it has been show that quantum Fisher information is equivalent to the metric of the Bures distance between two density
%operators up to an insignificant constant factor \cite{Ye2021}.
Quantum interferometric power (QIP) \cite{Girolami2014} is a measure of discord-type quantum correlation defined via quantum Fisher information, which is upper bounded by local quantum uncertainly (LQU) \cite{Girolami2013}. Actually, if subsystem ``$a$'' is a qubit, LQU is equivalent to Hellinger distance of discord \cite{Chang2013,roga2016}, and is upper bounded by Bures distance of discord. As a result, QIP is also upper bounded by Bures distance of discord in $C^2\otimes C^d$ case.

 %Actually, quantum Fisher information has also been used to characterize  and nonclassical correlations \cite{Kim2018}. Since QIP is bounded by local quantum uncertainty (LQU) \cite{Girolami2013}, which is equivalent to the Hellinger distance of discord  if subsystem a is a qubit, then the relationship between Hellinger distance and Bures distance implies that QIP is also upper bounded by $D^a_B$ in this case.
\end{rem}

\section{Characterizing entanglement via asymmetric quantum discord}
\label{characterization-asymmetric}
In this section, we propose the framework to quantify entanglement via quantum discord (asymmetric or local), and introduce several quantifiers with geometric discord and measurement-induced geometric discord.

\subsection{General results}
In this section, we explore and interpret the relationship between entanglement and quantum discord (asymmetric) over state extensions.

\begin{figure}%
%\centering
%\resizebox{0.6\columnwidth}{!}{
\includegraphics[width = 3in]{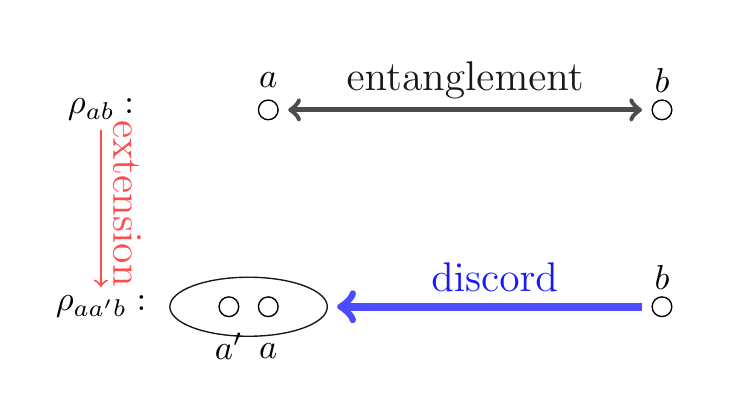}%
%}%
\caption{The minimal discord of extended state $\rho_{aa^{\prime}b}$ between parties $aa^{\prime}$ and $b$ is a quantification of entanglement of $\rho_{ab}$ between $a$ and $b$.}
\label{fig1}
\end{figure}

\begin{defn}
For a bipartite state $\rho_{ab}\in D(\mathcal{H})$, the minimal discord over state extensions is defined as
\begin{align}
\hat{\mathcal{E}}(\rho_{ab}):=\min_{\textmd{tr}_{a^{\prime}}[\rho_{aa^{\prime}b}]=\rho_{ab}}\hat{D}(\rho_{aa^{\prime}b}),
\end{align}
where the minimization is taken over all state extensions $\rho_{aa^{\prime}b}$ of $\rho_{ab}$ \cite{note-state-extension} along the bipartition $aa^{\prime}:b$ (see Fig. \ref{fig1}). 
%{\color{red} Obviously, the purified states of $\rho_{ab}$ form a subset of its extended states.}
\end{defn}

\begin{rem}
Squashed entanglement \cite{christandl2004} and the conditional entanglement of mutual information \cite{Yang2008}, two additive entanglement measures, are defined using the notion of ``over state extensions". Thus, characterizing quantum entanglement using extended systems has proved an important idea in entanglement theory.
\end{rem}

\begin{thm}\label{theorem-ent-asymmetric-discord}
If $\hat{D}$ is a discord measure satisfying (D1-D4), the corresponding minimal discord, $\hat{\mathcal{E}}$, over state extensions has the following remarkable properties:

\noindent(E1) $\hat{\mathcal{E}}(\rho_{ab})\ge0$ with the equality if and only if $\rho\in\mathcal{S}$.

\noindent(E2) $\hat{\mathcal{E}}$ is invariant under local unitary transformations.

\noindent(E3) $\hat{\mathcal{E}}$ is nonincreasing under local partial trace,
\begin{align*}
\hat{\mathcal{E}}(\rho_{ab})\le\hat{\mathcal{E}}(\rho_{aa_1b})
\end{align*}
for any state extension $\rho_{aa_1b}$ of $\rho_{ab}$.

\noindent(E4) $\hat{\mathcal{E}}$ is nonincreasing under local operations in party \(a\).

\noindent(E5) $\hat{\mathcal{E}}$ reduces to an entanglement monotone for pure states.
\end{thm}

\begin{proof}
\noindent(E1) The nonnegativity of quantum discord implies that $\hat{\mathcal{E}}$ is always nonnegative. Moreover, any separable state $\rho_{ab}=\sum_ip_i\rho^i_a\otimes\rho^i_b$ can be embedded into a larger classical-quantum state $\rho_{aa^{\prime}b}=\sum_ip_i\ket{\alpha_i}_{aa^{\prime}}\bra{\alpha_i}\otimes\rho^i_b$ such that $\rho_{ab}=tr_{a^{\prime}}[\rho_{aa^{\prime}b}]$, where $a^{\prime}$ is the ancillary system pertinent to party \(a\) and $\ket{\alpha_i}_{aa^{\prime}}$ is the purification of $\rho^i_a$ for each $i$ \cite{devi2008}. However, an entangled state does not admit such an extension. % Consequently, a bipartite state is separable if and only if it is a reduced state of a classical quantum state.
Therefore, $\hat{\mathcal{E}}$ is faithful in separable states.

\noindent(E2) Assuming $\rho_{aa^{\prime}b}$ is the state extension of $\rho_{ab}$, $tr_{a^{\prime}}[U_a\otimes U_b\rho_{aa^{\prime}b}U^{\dagger}_a\otimes U^{\dagger}_b]=U_a\otimes U_b\rho_{ab}U^{\dagger}_a\otimes U^{\dagger}_b$ and the local unitary invariance of $\hat{D}$ implies that $\hat{\mathcal{E}}(\rho_{ab})\ge\hat{\mathcal{E}}(U_a\otimes U_b\rho_{ab}U^{\dagger}_a\otimes U^{\dagger}_b)$. On the contrary, we can also show $\hat{\mathcal{E}}(\rho_{ab})\le\hat{\mathcal{E}}(U_a\otimes U_b\rho_{ab}U^{\dagger}_a\otimes U^{\dagger}_b)$ implying that $\hat{\mathcal{E}}$ is invariant under local unitary transformation.

\noindent(E3) This follows trivially because any state extension $\rho_{aa_1a^{\prime}b}$ of state $\rho_{aa_1b}$ is also the extension of $\rho_{ab}$.

\noindent(E4) % $\mathcal{E}^a$ is non-increasing under local operations in party a. %Since $D^a$ is non-increasing in the local operation in subsystem b, then it is enough to prove subsystem a.
Using Stinespring representation \cite{stinespring1955}, the local operation in party \(a\) can be realized by adding a pure state ancilla, followed by a global unitary operation and tracing out the ancilla system, i.e., $\sum_iK^a_i\rho_{ab}K^{a\dagger}_i=tr_{a_1}U_{aa_1}(\rho_{ab}\otimes\ket{0}_{a_1}\bra{0})U^{\dagger}_{aa_1}$. Therefore, one has
\begin{align}
\hat{\mathcal{E}}(\rho_{ab})&\ge\hat{\mathcal{E}}(\rho_{ab}\otimes\ket{0}_{a_1}\bra{0}) \nonumber \\
&=\hat{\mathcal{E}}(U_{aa_1}\rho_{ab}\otimes\ket{0}_{a_1}\bra{0}U^{\dagger}_{aa_1})\nonumber \\
&\ge \hat{\mathcal{E}}(\sum_iK^a_i\rho_{ab}K^{a\dagger}_i),
\end{align}
where the inequality in the first and the third lines follows from property (E3).
%Note that the state in second line is an extension of $\sum_iK^a_i\rho_{ab}K^{a\dagger}_i$, then property 3 implies the last inequality.

\noindent(E5) %$\mathcal{E}^a$ reduces to the entanglement measure for pure states.
Firstly, combining the definition of $\hat{\mathcal{E}}$ and property (E3),
\begin{align}
\hat{\mathcal{E}}(\ket{\psi})\ge\min_{\ket{\phi}}\hat{D}(\ket{\psi}\otimes\ket{\phi})\ge \hat{D}(\ket{\psi})=E(\ket{\psi}).
\end{align}
Further, considering the special case, that is, the extension space is one-dimensional, $\hat{\mathcal{E}}(\ket{\psi})=\hat{D}(\ket{\psi})=E(\ket{\psi})$.
\end{proof}

\begin{rem}
The above results provide an alternative avenue to understand quantum entanglement from the viewpoint of discord over state extensions. We can see that $\hat{\mathcal{E}}$ is a good candidate of an entanglement measure. Moreover, $\hat{D}$ reduces to $\hat{\mathcal{E}}$ for pure states.
$\hfill \square$
\end{rem}

We call a nonnegative bivariate function $d$ on state space {\it pseudo-distance} if $d(\rho,\sigma)=0$ iff $\rho=\sigma$, and 
call $d$ contractive, if it satisfies $d(\rho,\sigma) \ge d(\Phi(\rho),\Phi(\sigma))$ for any quantum operation $\Phi$ and $\rho,\sigma\in D(\mathcal{H})$. Relative entropy and Bures distance are examples of pseudo-distances. We will simply write ``distance'' for simplicity.
%We remark that by {\it pseudo-distance}, we refer to a nonnegative bivariate function $d$ in state space with $d(\rho,\sigma)=0$ iff $\rho=\sigma$.

%Next, we consider the geometric discord and measurement induced geometric discord case.

\begin{defn}
For a bipartite state $\rho_{ab}\in D(\mathcal{H})$, the minimal geometric discord over state extensions (GDSE) is defined as
\begin{align}
\label{eq-gdse-def}
\hat{\mathcal{E}}_d(\rho_{ab}):=\min_{\sigma_{aa^{\prime}b}\in\mathcal{CQ}}\min_{\textmd{tr}_{a^{\prime}}[\rho_{aa^{\prime}b}]=\rho_{ab}}d(\rho_{aa^{\prime}b},\sigma_{aa^{\prime}b}),
\end{align}
where the minimization is taken over all extended states $\rho_{aa^{\prime}b}$ and classical-quantum states in $D(\mathcal{H}_{aa^{\prime}b})$. Here $d$ is a contractive distance in $D(\mathcal{H}_{aa^{\prime}b})$. Furthermore,
the minimal measurement-induced geometric discord over state extensions (MIDSE) is defined as
\begin{align}
\label{eq-midse-def}
\hat{\mathcal{E}}^{\prime }_d(\rho_{ab}):=\min_{\Pi_{aa^{\prime}}}\min_{\textmd{tr}_{a^{\prime}}[\rho_{aa^{\prime}b}]=\rho_{ab}}d(\rho_{aa^{\prime}b},\Pi_{aa^{\prime}}\otimes I_b(\rho_{aa^{\prime}b}))
\end{align}
where the minimization is taken over extended states and local projection in subsystem $aa^{\prime}$.% and respect to bipartite $aa^{\prime}:b$ .
\end{defn}

\begin{rem}
For both relative entropy and Bures distance, the corresponding quantifiers of discord satisfy (D1-D4). Theorem \ref{theorem-ent-asymmetric-discord} then implies that the corresponding GDSE and MIDSE are good candidates of entanglement measures.
$\hfill \square$
\end{rem}

Next, we consider the quantification of entanglement by performing partial trace; a slight modification of Eq. (\ref{eq-gdse-def}). Define quantum correlation by
\begin{align}
\check{\mathcal{E}}_d(\rho_{ab}):=\min_{\sigma_{aa^{\prime}b}\in\mathcal{CQ}}  d(\rho_{ab},tr_{a^{\prime}}\sigma_{aa^{\prime}b}),
\end{align}
%\min_{\textmd{tr}_{a^{\prime}}[\rho_{aa^{\prime}b}]=\rho_{ab}}
where the minimum is taken over all
%extended states $\rho_{aa^{\prime}b}$ and
classical-quantum state in $D(\mathcal{H}_{aa^{\prime}b})$. Obviously, $\check{\mathcal{E}}_d$ is equivalent to $E_d:=\min_{\sigma_{ab}\in\mathcal{S}}d(\rho_{ab},\sigma_{ab})$, which is the corresponding entanglement quantification of distance $d$. In other words, geometric entanglement can be linked to discord over state extensions. For measurement-induced geometric discord, however, it is not the trivial case, as shown below.

\begin{defn}
For $\rho_{ab}\in D(\mathcal{H}_{ab})$, we define a quantity
\begin{align}
\check{\mathcal{E}}^{\prime }_d(\rho_{ab}):=\min_{\Pi_{aa^{\prime}}}\min_{\textmd{tr}_{a^{\prime}}[\rho_{aa^{\prime}b}]=\rho_{ab}}d(\rho_{ab},tr_{a^{\prime}}[\Pi_{aa^{\prime}}\otimes I_b(\rho_{aa^{\prime}b})]),
\end{align}
where the minimal is taken over all  extended states $\rho_{aa^{\prime}b}$ and local projection in subsystem $aa^{\prime}$. %and the symbol $\mathcal{E}^{\prime}$  represents ``measurement induced discord''.
\end{defn}

This quantity is related to the previous ones, via the following inequalities.
\begin{thm} \label{thm-inequalities}
For $\rho_{ab}\in D(\mathcal{H}_{ab})$,
\begin{align}
\hat{\mathcal{E}}^{\prime}_d(\rho_{ab})\ge\hat{\mathcal{E}}_d(\rho_{ab})\ge E_d(\rho_{ab}), \\
\check{\mathcal{E}}^{\prime }_d(\rho_{ab})\ge\check{\mathcal{E}}_d(\rho_{ab})=E_d(\rho_{ab}),
\end{align}
and
\begin{align}
&\hat{\mathcal{E}}^{\prime}_d(\rho_{ab})\ge\check{\mathcal{E}}^{\prime a}_d(\rho_{ab}), \\ &\hat{\mathcal{E}}_d(\rho_{ab})\ge\check{\mathcal{E}}_d(\rho_{ab})=E_d(\rho_{ab}).
\end{align}
\end{thm}
\begin{proof}
These inequalities can be derived from the definition directly.
\end{proof}

%%%
\subsection{Characterization with Bures distance discord}
Here, we prove two important theorems related to the minimal Bures distance discord, $\hat{\mathcal{E}}_B$, over state extensions. In the following theorem we show that $\mathcal{E}^a_B$ is convex.

\begin{thm}\label{asymmetric-convex-Bures-discord}
$\hat{\mathcal{E}}_B$ is convex,
\begin{align}
\hat{\mathcal{E}}_B(\sum_ip_i\rho^i_{ab})\le\sum_ip_i\hat{\mathcal{E}}_B(\rho^i_{ab}),
\end{align}
where $p_i$ are probabilities and $\rho^i_{ab}$ are bipartite states shared
between parties $a$ and $b$.
\end{thm}

\begin{proof}
See Appendix \ref{proof-theorem-asymmetric-convex}.
\end{proof}

Using Theorem \ref{asymmetric-convex-Bures-discord}, we arrive at another theorem below.
\begin{thm}\label{main-theorem}
For $\rho_{ab}\in D(\mathcal{H})$, the minimal Bures distance of discord over state extensions is equivalent to the Bures distance of entanglement,
\begin{align}
\hat{\mathcal{E}}_B(\rho_{ab})=E_B(\rho_{ab}).
\end{align}
\end{thm}

\begin{proof}
%\begin{rem}
Since $E_B(\rho_{ab})=E^{cr}_B(\rho_{ab})$ \cite{streltsov2010}, we just need to show that for each $\rho_{ab}\in D(\mathcal{H})$, we have
\begin{align}
E_B(\rho_{ab}) \le \hat{\mathcal{E}}_B(\rho_{ab}) \le E^{cr}_B(\rho_{ab}).
\end{align}
%\end{rem}
See Appendix \ref{proof-main-theorem} for the complete proof.
\end{proof}

Having established these results, Theorem \ref{thm-inequalities} for the Bures distance together with Theorem \ref{main-theorem} imply
%\begin{align*}
%&\mathcal{E}^{\prime a}_B(\rho_{ab})\ge\mathcal{E}^a_B(\rho_{ab})= E_B(\rho_{ab}),\\
%&\hat{\mathcal{E}}^{\prime a}_B(\rho_{ab})\ge\hat{\mathcal{E}}^{a}_B(\rho_{ab})=E_B(\rho_{ab}).
%\end{align*}
%and
\begin{align}
\hat{\mathcal{E}}^{\prime}_B(\rho_{ab})\ge\check{\mathcal{E}}^{\prime}_B(\rho_{ab})\ge E_B(\rho_{ab})=\hat{\mathcal{E}}_B(\rho_{ab})=\check{\mathcal{E}}_B(\rho_{ab}).
\end{align}

\subsection{Operational meaning of Bures distance of entanglement}

The quest for an operational meaning or interpretation of an entanglement measure lies at the very heart of the entanglement theory.
While entanglement of formation, distillable entanglement or entanglement cost are measures of entanglement having an operational meaning \cite{horodeckiRMP09,nielsen10}, the Bures distance of entanglement is an entanglement measure defined from the geometric viewpoint and its physical meaning is not clear. By means of the equivalence in Theorem \ref{main-theorem}, and the operational meaning of Bures distance of discord \cite{spehner2013a}, we provide an operational meaning of the Bures distance of entanglement.

Let us briefly review the ambiguous quantum state discrimination (QSD) protocol \cite{bergou2004}. Suppose Alice chooses a state $\rho_i$ from a set of states $\{\rho_i\}$ with probability $\eta_i$ and sends it to Bob, who determines which state he receives by performing a positive-perator-valued measure (POVM). %Since
% on each $\rho_i$, and declares that the state is $\rho_j$ when the measurement outcome reads $j$. %The POVM is a set of positive operators $\{M_i\}$ satisfying $ \sum_iM_i =I$.
Since the probability to get the result $j$ with given state $\rho_i$ is $p(j|i)=tr(M_j\rho_i)$, then the corresponding optimal success probability is
\begin{align}
P^{opt}_s(\{\rho_i,\eta_i\}^N_{i=1}):=\max_{M_i}\sum_i\eta_itr(M_i\rho_i),
\end{align}
where the maximization is done over all POVMs. Similarly, $P^{opt~(vN)}_s(\{\rho_i,\eta_i\}^N_{i=1})=\max_{\Pi_i}\sum_i\eta_itr(\Pi_i\rho_i)$ is the optimal success probability to discriminate $\{\rho_i,\eta_i\}^N_{i=1}$ with von Neumann measurement.

Based on the operational meaning of the Bures distance of asymmetric discord \cite{spehner2013a}, %where the zero-discord state is of the form $\sum_ip_i\ket{\alpha_i}\bra{\alpha_i}\otimes\rho_i$,
the connection between geometric entanglement and QSD is as follows. %(see in Appendix.\ref{discord and QSD}).

\begin{cor}
For a bipartite quantum state $\rho_{ab}\in D(\mathcal{H})$, the square of the maximal fidelity to the set of separable states, $F(\rho_{ab},\mathcal{S}):=\max_{\sigma_{ab} \in \mathcal{S}}F(\rho_{ab},\sigma_{ab})$, is equal to the optimal success probability to discriminate a set of quantum states with von Neumann measurement, i.e.,
\begin{align}\label{operational-meaning-Bures-discord}
F^2(\rho_{ab},\mathcal{S})=\max_{\rho_{aa^{\prime}b},\ket{\alpha_i}}P^{opt~(vN)}_s(\{\rho_i,\eta_i\}),
\end{align}
where $\eta_i=tr\bra{\alpha_i}\rho_{aa^{\prime}b}\ket{\alpha_i}$, $\rho_i=\eta^{-1}_i\sqrt{\rho_{aa^{\prime}b}}\ket{\alpha_i}\bra{\alpha_i}$ $\sqrt{\rho_{aa^{\prime}b}}$ and the maximum is taken over all possible extended states $\rho_{aa^{\prime}b}$, von Neumann measurement $\Pi_i$ and orthogonal basis $\{\ket{\alpha_i}\}$ on $\mathcal{H}_{aa^{\prime}}$.
\end{cor}
This offers an operational meaning to the Bures distance of entanglement.

\subsection{Characterization with relative entropy discord}

We define relative entropy of discord by $\hat{D}_r(\rho):=\min_{\sigma\in\mathcal{CQ}} S(\rho||\sigma)$ and measurement-induced relative entropy of discord by $\hat{D}^{\prime}_r(\rho):=\min_{\Pi_a} S(\rho||\Pi_a(\rho))$. %where the minimal over all local von Neumann measurement $\Pi_a$ in subsystem a.

\begin{rem}
In Ref. \cite{baumgratz2014} the authors provided a framework to quantify quantum coherence with respect to a chosen orthogonal basis. Although discord and coherence seem different, they have essential similarities.
Discord can be regarded as the basis-independent coherence \cite{Yao2015}, and relative entropy of discord is equivalent to the minimal relative entropy of partial coherence \cite{sun2017}. Moreover, quantum discord can also be explained as an upper bound to the quantum correlations generated from partial coherence via partial incoherent operations \cite{Kim2018b}.
\end{rem}

In the following theorem, we show that $\hat{D}_r$ and $\hat{D}^{\prime}_r$ are equivalent.

\begin{thm}\label{lemma-relative-entropy}
For a bipartite state $\rho_{ab}\in D(\mathcal{H})$, the relative entropy of discord is equivalent to the measurement-induced relative entropy of discord, i.e.,
\begin{align}
\hat{D}_r(\rho_{ab})=\hat{D}^{\prime}_r(\rho_{ab}).
\end{align}
\end{thm}

\begin{proof}
%For a classical-quantum state $\sigma_{cq}=\sum_jp_j\pi^j_a\otimes\sigma_j$, one has that
%\begin{align*}
%tr(\rho_{ab}\log\sigma_{cq})%&=\sum_j\log p_jtr(\rho_{ab}\pi^j_a\otimes\sigma_j)\\
%%&=\sum_j\log p_jtr[\Pi_a(\rho_{ab})\pi^j_a\otimes\sigma_j]\\
%&=tr[\Pi_a(\rho_{ab})\log\sigma_{cq}],
%\end{align*}
%where $\Pi_a(\rho_{ab})=\sum_i\pi^i_a\otimes I_b(\rho_{ab})$. Then,
%\begin{align*}
%S(\rho||\sigma_{cq})&=tr[\rho\log\rho]-tr[\rho\log\sigma_{cq}]\\
%&=tr[\rho\log\rho]-tr[\Pi_a(\rho_{ab})\log\Pi_a(\rho_{ab})]\\
%&+tr[\Pi_a(\rho_{ab})\log\Pi_a(\rho_{ab})]-tr[\Pi_a[\rho_{ab}]\log\sigma_{cq}]\\
%&=S(\Pi_a[\rho_{ab}])-S(\rho_{ab})+S(\Pi_a[\rho_{ab}]||\sigma_{cq}).
%\end{align*}

Since any classical-quantum state has the form $\sigma_{ab}^{(cq)}=\sum_ip_i\ket{i}_a\bra{i}\otimes\sigma^i_{b}$, using Eq. (\ref{eq-relent}) we have
\begin{align}
\hat{D}_r(\rho_{ab})&=\min_{\pi^i_a}\min_{p_i,\sigma^i_b}S(\rho_{ab}||\sigma_{ab}^{(cq)}) \nonumber \\
&=\min_{\pi^i_a} [S(\Pi_a (\rho_{ab}))-S(\rho_{ab})+\min_{p_i,\sigma^i_b} S(\Pi_a (\rho_{ab})||\sigma_{ab}^{(cq)})] \nonumber \\
&=\min_{\pi^i_a} S(\Pi_a (\rho_{ab}))-S(\rho_{ab}) \nonumber \\
&=\min_{\pi^i_a}S(\rho_{ab}||\Pi_a (\rho_{ab}))=\hat{D}^{\prime}_r(\rho_{ab}),
\end{align}
where $\Pi_a(\rho_{ab})=\sum_i(\ket{i}_a\bra{i}\otimes I_b)\rho_{ab}(\ket{i}_a\bra{i}\otimes I_b)$.
\end{proof}

Next, for the relative entropic quantities $E_r$, $\hat{\mathcal{E}}^{\prime}_r$, $\hat{\mathcal{E}}_r$, $\check{\mathcal{E}}^{\prime}_r$ and $\check{\mathcal{E}}_r$, where symbols have their usual meanings, Theorem \ref{thm-inequalities} and Theorem \ref{lemma-relative-entropy} lead to the following result.
\begin{thm}
For a bipartite state $\rho_{ab}\in D(\mathcal{H})$,
\begin{align}
\hat{\mathcal{E}}^{\prime}_r(\rho_{ab})=\hat{\mathcal{E}}_r(\rho_{ab})\ge E_r(\rho_{ab})=\check{\mathcal{E}}^{\prime}_r(\rho_{ab})=\check{\mathcal{E}}_r(\rho_{ab}).
\end{align}
\end{thm}

\begin{proof}
From Theorem \ref{lemma-relative-entropy}, we have
\begin{align}
\hat{\mathcal{E}}^{\prime}_r(\rho_{ab})=\hat{\mathcal{E}}_r(\rho_{ab}),
\end{align}
and
\begin{align}
\check{\mathcal{E}}^{\prime}_r(\rho_{ab})=\check{\mathcal{E}}_r(\rho_{ab})=E_r(\rho_{ab}).
\end{align}
The inequality follows from Theorem \ref{thm-inequalities}.
\end{proof}

\begin{rem}\label{remark-relent}
In Ref. \cite{devi2008}, authors proved that
$
 \check{\mathcal{E}}^{\prime}_r(\rho_{ab})=\min_{tr_a[\sigma^{(sep)}_{ab}]=\rho_b}S(\rho_{ab}||\sigma^{(sep)}_{ab})\ge E_r(\rho_{ab}),
$
where the minimum is taken over all separable states $\sigma^{(sep)}_{ab}$ with $tr_a[\sigma^{(sep)}_{ab}]=\rho_b$.
%Obviously, our results implies that the $``="$ always holds.
Our results affirm that they are equal.
$\hfill \square$
\end{rem}

\section{Characterizing entanglement via symmetric discord}
\label{characterization-symmetric}
In this section, we quantify entanglement via symmetric or global quantum discord.
%and introduce several quantifiers with geometric discord and measurement-induced geometric discord.
%
In the symmetric or global quantum discord of a bipartite system, the local measurement is performed on both subsystems.
%We ignore ``symmetric'' and still call it discord without ambiguity.
Without further ado, all the contents developed in the asymmetric case will be recalled here, with the only difference that now quantum measurement or other treatment will be on both parties.

\subsection{General results}

\begin{figure}%
%\centering
%\resizebox{0.6\columnwidth}{!}{
\includegraphics[width = 3in]{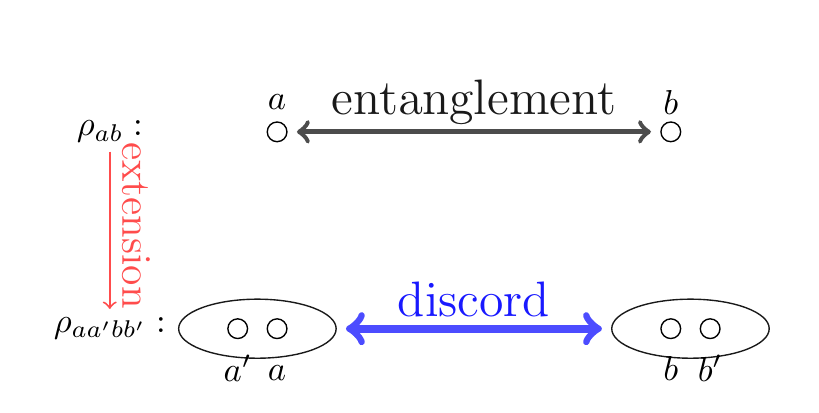}%
%}%
\caption{The minimal discord of an extended state $\rho_{aa^{\prime}bb^{\prime}}$ shared between parties $aa^{\prime}$ and $bb^{\prime}$ is a quantification of entanglement of $\rho_{ab}$ between $a$ and $b$.}
\label{fig2}
\end{figure}

Following, we define the minimal quantum discord over state extensions. In the asymmetric case, the state extension was in party \(a\). Here we consider state extension in both parties.
\begin{defn}
For bipartite state $\rho_{ab}\in D(\mathcal{H})$, the minimal discord over state extensions is defined as
\begin{align}
\tilde{\mathcal{E}}(\rho_{ab}):=\min_{\textmd{tr}_{a^{\prime}b^{\prime}}[\rho_{aa^{\prime}bb^{\prime}}]=\rho_{ab}}\tilde{D}(\rho_{aa^{\prime}bb^{\prime}}),
\end{align}
where the minimization is taken along the bipartition $aa^{\prime}:bb^{\prime}$ with $tr_{a^{\prime}}\rho_{aa^{\prime}bb^{\prime}}=\rho_{ab}$ (see Fig. \ref{fig2}).
\end{defn}

\begin{thm}\label{theorem-ent-symmetric-discord}
If $\tilde{D}$ is a symmetric discord measure satisfying (D${'}$1-D${'}$3), the minimal discord, $\tilde{\mathcal{E}}$, over state extensions has the following desirable and remarkable properties:

\noindent(E$^{\prime}$1) $\tilde{\mathcal{E}}(\rho_{ab})\ge0$ with the equality if and only if $\rho\in\mathcal{CC}$.

\noindent(E$^{\prime}$2) $\tilde{\mathcal{E}}$ is invariant under local unitary transformations.

\noindent(E$^{\prime}$3) $\tilde{\mathcal{E}}$ is nonincreasing under local partial trace,
\begin{align}
\tilde{\mathcal{E}}(\rho_{ab})\le\tilde{\mathcal{E}}(\rho_{aa_1bb_1}),
\end{align}
 for any state extension $\rho_{aa_1bb_1}$ of $\rho_{ab}$.

\noindent(E$^{\prime}$4) $\tilde{\mathcal{E}}$ is nonincreasing under local operations.

\noindent(E$^{\prime}$5) $\tilde{\mathcal{E}}$ reduces to an entanglement monotone for pure states.
\end{thm}

\begin{proof}
The proof is similar to that of Theorem \ref{theorem-ent-asymmetric-discord}.
\end{proof}

\subsection{Minimal Bures distance of symmetric discord}

%In \cite{Luo2016}, Luo proposed that the minimal symmetric discord of bipartite quantum state over state extensions can be used to characterize entanglement.

In this part, we define the minimal Bures distance of symmetric discord.

\begin{defn}
For a bipartite state $\rho_{ab}\in\mathcal{H}$, the minimal Bures distance of discord over state extensions is defined as
\begin{align}
\tilde{\mathcal{E}}_B(\rho_{ab}):=\min_{\sigma\in\mathcal{CC}}\min_{\textmd{tr}_{a^{\prime}b^{\prime}}[\rho_{aa^{\prime}bb^{\prime}}]=\rho_{ab}}d^2_B(\rho_{aa^{\prime}bb^{\prime}},\sigma),
\end{align}
where the minimum is taken with respect to the bipartition $aa^{\prime}:bb^{\prime}$.
\end{defn}

Following, we state that the minimal Bures distance of symmetric discord is convex, and is equivalent to the Bures distance of entanglement.

\begin{thm}
$\tilde{\mathcal{E}}_B$ is convex,
\begin{align}
\tilde{\mathcal{E}}_B(\sum_ip_i\rho^i_{ab})\le\sum_ip_i\tilde{\mathcal{E}}_B(\rho^i_{ab}),
\end{align}
where $p_i$ are probabilities and $\rho^i_{ab}$ are bipartite states shared
between parties $a$ and $b$.
\end{thm}
The proof is similar to that of Theorem \ref{asymmetric-convex-Bures-discord}.
With above Theorem, one has the following result.
\begin{thm}\label{main-theorem-1}
For $\rho_{ab}\in D(\mathcal{H})$,
%\begin{align*}
$\tilde{\mathcal{E}}_B(\rho_{ab})=E_B(\rho_{ab}).$
%\end{align*}
\end{thm}
The proof is similar to that of Theorem \ref{main-theorem}.

\begin{rem}
It is interesting to note that $E_B(\rho_{ab})=\hat{\mathcal{E}}_B(\rho_{ab})=\tilde{\mathcal{E}}_B(\rho_{ab})$, which means that Bures distance of entanglement is equivalent to the minimal Bures distance of discord over state extensions, both on one subsystem and two subsystems.
$\hfill \square$
\end{rem}
%On one hand, this equivalence links discord to entanglement, which also offer another interesting perspective on entanglement, that is, entanglement can be regards as the irreducible part of quantum discord from the perspective of a small system. On the second hand, it is inconceivable that Bures distance of discord over one side state extension is equivalent to two side case.

\begin{rem}
It was Luo who first proposed to quantify entanglement as the minimal quantum discord over state extensions \cite{luo2016}. Above theorem offers an affirmative evidence that this kind of entanglement quantification is consistent with the previous entanglement measures. However, it is not clear whether this equivalence still holds for other entanglement measures such as entanglement measures based on distances \cite{horodeckiRMP09} and relative entropy of entanglement \cite{vedral1997}.
$\hfill \square$
\end{rem}

\subsection{More definitions}

%\begin{defn}
%For bipartite state $\rho_{ab}\in\mathcal{H}$, the minimal discord over state extensions is defined as
%\begin{align*}
%\mathcal{E}^a(\rho_{ab}):=\min_{tr_{a^{\prime}b^{\prime}}\rho_{aa^{\prime}bb^{\prime}}=\rho_{ab}}D^a(\rho_{aa^{\prime}bb^{\prime}}),
%\end{align*}
%where the minimal discord is taken over bipartite $aa^{\prime}:bb^{\prime}$ with %$tr_{a^{\prime}b^{\prime}}\rho_{aa^{\prime}bb^{\prime}}=\rho_{ab}$.
%\end{defn}

\begin{defn}
For a bipartite state $\rho_{ab}\in D(\mathcal{H})$, the minimal geometric discord over state extensions (GDSE) is defined as
\begin{align}
\tilde{\mathcal{E}}_d(\rho_{ab}):=\min_{\sigma_{aa^{\prime}bb^{\prime}}\in\mathcal{CC}}\min_{\textmd{tr}_{a^{\prime}b^{\prime}}[\rho_{aa^{\prime}bb^{\prime}}]=\rho_{ab}}d(\rho_{aa^{\prime}bb^{\prime}},\sigma_{aa^{\prime}bb^{\prime}}),
\end{align}
where the minimum is taken with respect to the bipartition $aa^{\prime}:bb^{\prime}$. Moreover,
the minimal measurement-induced geometric discord over state extensions (MIDSE) is defined as
\begin{align}
\resizebox{0.96\hsize}{!}{%
$\tilde{\mathcal{E}}^{\prime }_d(\rho_{ab}):=\min_{\Pi_{aa^{\prime}}, \Pi_{bb^{\prime}}} \min_{\textmd{tr}_{a^{\prime}b^{\prime}}[\rho_{aa^{\prime}bb^{\prime}}]=\rho_{ab}}d(\rho_{aa^{\prime}bb^{\prime}},\Pi_{aa^{\prime}}\otimes \Pi_{bb^{\prime}}(\rho_{aa^{\prime}bb^{\prime}}))$,%
}
\end{align}
where the minimum is taken over all local projections in both subsystems $aa^{\prime}~\&~ bb^{\prime}$ and along the bipartition $aa^{\prime}:bb^{\prime}$.
%{\color{red}Comment: Here, the minimization should be over all local projections in both subsystems $aa^{\prime}$ and $bb^{\prime}$, right? This is the symmetric case.}
\end{defn}

\begin{rem}
For relative entropy or Bures distance, the corresponding geometric discord satisfies (D$^{\prime}$1-$D^{\prime}$3). From Theorem \ref{theorem-ent-symmetric-discord}, the corresponding quantification is a good candidate of entanglement measure.
$\hfill \square$
\end{rem}

Next, let us consider the quantification by performing partial trace,
\begin{align}
\breve{\mathcal{E}}_d(\rho_{ab}):=\min_{\sigma_{aa^{\prime}bb^{\prime}}\in\mathcal{CC}}  d(\rho_{ab},tr_{a^{\prime}b^{\prime}}\sigma_{aa^{\prime}bb^{\prime}}),
\end{align}
%\min_{\textmd{tr}_{a^{\prime}b^{\prime}}[\rho_{aa^{\prime}bb^{\prime}}]=\rho_{ab}}
where the minimum is taken over all
%extended states $\rho_{aa^{\prime}bb^{\prime}}$ and corresponding
classical-classical states. It is easy to see that $\check{\mathcal{E}}_d$ is equivalent to the general geometric entanglement measure $E_d$, which is defined as the minimal distance between the state and the set of separable states.
Nevertheless, it is not the case for the measurement-induced geometric discord, as shown below.

\begin{defn}
For $\rho_{ab}\in D(\mathcal{H}_{ab})$, we define a quantity $\breve{\mathcal{E}}^{\prime}_d(\rho_{ab})$ $:=\min_{\Pi_{aa^{\prime}},\Pi_{bb^{\prime}}} \min_{\textmd{tr}_{a^{\prime}b^{\prime}}[\rho_{aa^{\prime}bb^{\prime}}]=\rho_{ab}} $ $d(\rho_{ab},$ $tr_{a^{\prime}b^{\prime}}$ $[\Pi_{aa^{\prime}}\otimes \Pi_{bb^{\prime}}(\rho_{aa^{\prime}bb^{\prime}})]),$
%\end{align*}
where the minimum is taken over all extended states $\rho_{aa^{\prime}bb^{\prime}}$ and local projection in both subsystems $aa^{\prime}$ and $bb^{\prime}$.
%{\color{red}Comment: Here, the minimization should be over all local projections in both subsystems $aa^{\prime}$ and $bb^{\prime}$, right? This is the symmetric case.}
\end{defn}

%Obviously, any separable state $\rho_{ab}$ has a separable extended state and there exist an orthogonal local projective measurement $\Pi_{aa^{\prime}},\Pi_{bb^{\prime}}$ which keeps the later untouched, then $\hat{\mathcal{E}}^{\prime}_d(\rho_{ab})=0$. % Moreover, since the local partial trace

Certainly, we obtain $\breve{\mathcal{E}}^{\prime}_d(\rho_{ab})\ge E_d(\rho_{ab})$ for any $\rho_{ab}\in D(\mathcal{H}_{ab})$. %because $tr_{a^{\prime}b^{\prime}}[\Pi_{aa^{\prime}}\otimes \Pi_{bb^{\prime}}(\rho_{aa^{\prime}bb^{\prime}})]$ is a separable state.
Moreover, based on the discussion in Remark \ref{remark-relent}, the equality holds for relative entropy. And whether the equality holds for other distances as well would be an interesting investigation.

In conclusion, we have the following result.
\begin{thm}
For $\rho_{ab}\in D(\mathcal{H}_{ab})$, the following inequalities are true.
\begin{align}
\tilde{\mathcal{E}}^{\prime}_d(\rho_{ab})\ge\tilde{\mathcal{E}}_d(\rho_{ab})\ge E_d(\rho_{ab}), \\ \breve{\mathcal{E}}^{\prime}_d(\rho_{ab})\ge\breve{\mathcal{E}}_d(\rho_{ab})=E_d(\rho_{ab}),
\end{align}
and
\begin{align}
&\tilde{\mathcal{E}}^{\prime}_d(\rho_{ab})\ge\breve{\mathcal{E}}^{\prime }_d(\rho_{ab}), \\ &\tilde{\mathcal{E}}_d(\rho_{ab})\ge\breve{\mathcal{E}}_d(\rho_{ab})=E_d(\rho_{ab}).
\end{align}
\end{thm}
\begin{proof}
The inequalities can be derived directly from the definitions.
\end{proof}

\subsection{Characterization with Bures distance discord \\ and relative entropy discord}
For the Bures distance, Theorem \ref{main-theorem-1} implies
%\begin{align*}
%&\mathcal{E}^{\prime a}_B(\rho_{ab})\ge\mathcal{E}^a_B(\rho_{ab})= E_B(\rho_{ab}),\\
%&\hat{\mathcal{E}}^{\prime a}_B(\rho_{ab})\ge\hat{\mathcal{E}}^{a}_B(\rho_{ab})=E_B(\rho_{ab}).
%\end{align*}
%and
\begin{align}
\tilde{\mathcal{E}}^{\prime}_B(\rho_{ab})\ge\breve{\mathcal{E}}^{\prime}_B(\rho_{ab})\ge E_B(\rho_{ab})=\mathcal{E}_B(\rho_{ab})=\breve{\mathcal{E}}_B(\rho_{ab}).
\end{align}

and for relative entropy,
%\begin{align*}
%E_B(\rho_{ab})=
%&\mathcal{E}^{\prime}_B(\rho_{ab})\ge\mathcal{E}_B(\rho_{ab})= E_B(\rho_{ab}),\\
%&\hat{\mathcal{E}}^{\prime}_B(\rho_{ab})\ge\hat{\mathcal{E}}_B(\rho_{ab})=E_B(\rho_{ab}).
%\end{align*}
%and
%\begin{align*}
%E_B(\rho_{ab})=\mathcal{E}_B(\rho_{ab})=\hat{\mathcal{E}}_B(\rho_{ab})=\mathcal{E}^a_B(\rho_{ab})=\hat{\mathcal{E}}^a_B(\rho_{ab}).
%\end{align*}

%Similarly, for relative entropy, it holds that
\begin{align*}
\tilde{\mathcal{E}}^{\prime}_r(\rho_{ab})=\tilde{\mathcal{E}}_r(\rho_{ab})\ge E_r(\rho_{ab})=\breve{\mathcal{E}}^{\prime}_r(\rho_{ab})=\breve{\mathcal{E}}^{\prime a}_r(\rho_{ab}).
\end{align*}

\section{Conclusion}\label{conclusion}
In this paper, we have proposed a framework to quantify entanglement from the viewpoint of discord, which unifies previous works including \cite{linan2008,devi2008,luo2016}. Several quantifications of entanglement are introduced based on quantum discord over state extensions.

Especially, for the Bures distance, we prove that the minimal discord (both asymmetric and symmetric) over state extensions is equivalent to the Bures distance of entanglement, which not only establishes an equivalence between this kind of entanglement quantification with existing entanglement measure, and but also provides an operational meaning for the Bures distance of entanglement. % Besides, it is interesting that the asymmetric discord over state extension is equal to the symmetric counterpart.

Moreover, for relative entropy, by proving that the corresponding discord measure is equivalent to the measurement induced relative entropy of discord, we show that the MIDSE (both asymmetric and symmetric) is equivalent to the relative entropy of entanglement, which  reinforces the result in Ref. \cite{devi2008}.

\begin{acknowledgments}
This project is supported in part by the Postdoctoral Science Foundation of China
(2021M702864), the National Natural Science Foundation of China (Grants No.  61876195, No. 61572532, No. 12050410232, No. 12031004, and No. 61877054), the Natural Science Foundation of Guangdong Province of China (Grant No. 2017B030311011), Jiangxi Provincial Natural Science Foundation (Grant No. 20202BAB201001) and the Fundamental Research Foundation for the Central Universities (Project No.K20210337).
\end{acknowledgments}

%

%\newpage
\section{Appendix}\label{appendix}
%\appendix

\subsection{Proof of Theorem \ref{asymmetric-convex-Bures-discord}}\label{proof-theorem-asymmetric-convex}

\begin{proof}
Note that
\begin{align}
\rho_{aa^{\prime}a^{\prime\prime}b}:=\sum_ip_i\rho^i_{aa^{\prime}b}\otimes\ket{i}_{a^{\prime\prime}}\bra{i}
\end{align}
is a state extension of $\rho_{ab}=\sum_ip_i\rho^i_{ab}$ whenever $\rho^i_{aa^{\prime}b}$ is a state extension of $\rho^i_{ab}$ for all i. Without loss of generality, suppose $\rho^i_{aa^{\prime}b}$ is the optimal state extension of $\rho^i_{ab}$ for each $i$, and $\sigma^{\star}_i$ is the corresponding closest classical-quantum state. Then, we have
\begin{align}
&\sum_ip_i\mathcal{E}^a_B(\rho^i_{ab})=\sum_ip_id^2_B(\rho^i_{aa^{\prime}b},\sigma^{\star}_i)\\
=&\sum_ip_id^2_B(\rho^i_{aa^{\prime}b}\otimes\ket{i}_{a^{\prime\prime}}\bra{i},\sigma^{\star}_i\otimes\ket{i}_{a^{\prime\prime}}\bra{i})\\
\ge & d^2_B(\sum_ip_i\rho^i_{aa^{\prime}b}\otimes\ket{i}_{a^{\prime\prime}}\bra{i},\sum_ip_i\sigma^{\star}_i\otimes\ket{i}_{a^{\prime\prime}}\bra{i})\\
\ge&\hat{D}_B(\sum_ip_i\rho^i_{aa^{\prime}b}\otimes\ket{i}_{a^{\prime\prime}}\bra{i})\\
\ge&\hat{\mathcal{E}}_B(\sum_ip_i\rho^i_{ab}),
\end{align}
where the first inequality follows from the joint convexity of $d^2_B$ and the second inequality is based on the definition of $\hat{D}_B$ and the fact that $\sum_ip_i\sigma^{\star}_i\otimes\ket{i}_{a^{\prime\prime}}\bra{i}$ is a classical-quantum state along the bipartition $aa^{\prime}a^{\prime\prime}:b$. The last inequality follows because $\sum_ip_i\rho^i_{aa^{\prime}b}\otimes\ket{i}_{a^{\prime\prime}}\bra{i}$ is a state extension of $\sum_ip_i\rho^i_{ab}$.
\end{proof}

\subsection{Proof of Theorem \ref{main-theorem}}\label{proof-main-theorem}

Here we restate the claim to be proved. That is, for each $\rho_{ab}\in D(\mathcal{H})$,
\begin{align}
E_B(\rho_{ab}) \le \hat{\mathcal{E}}_B(\rho_{ab}) \le E^{cr}_B(\rho_{ab}).
\end{align}

\begin{proof}
Suppose $\rho^{\star}_{aa^{\prime}b}$ is the optimal state extensions of $\rho_{ab}$ and $\sigma^{\star}$ is the corresponding closest classical-quantum state. Then
\begin{align}
\hat{\mathcal{E}}_B(\rho_{ab})=d^2_B(\rho^{\star}_{aa^{\prime}b},\sigma^{\star})\ge d^2_B(\rho_{ab},tr_{a^{\prime}}\sigma^{\star})\ge E_B(\rho_{ab}).
\end{align}
The first $"\ge"$ is the result of the contractibility of Bures distance and the second $"\ge"$ is because $tr_{a^{\prime}}\sigma^{\star}$ is a separable state. In fact, if $\sigma^{\star}=\sum_ip_i\ket{\alpha_i}_{aa^{\prime}}\bra{\alpha_i}\otimes\rho_i$, where
$\ket{\alpha_i}_{aa^{\prime}}=\sum_k\lambda^i_k\ket{x^i_k}_a\ket{y^i_k}_{a^{\prime}}$,
then tracing out the subsystem $a^{\prime}$ will lead to a decomposition
of the form $tr_{a^{\prime}}\sigma^{\star}=\sum_{i,k}p_i\lambda^i_k\ket{x^i_k}_a\bra{x^i_k}\otimes\rho_i$, that must be a separable state.

In particular, let us consider the pure state case. Suppose $\ket{\psi}=\sum_i\sqrt{\lambda_i}\ket{x_i}_a\ket{y_i}_b$ with $\lambda_1\ge...\ge\lambda_n$. Then
\begin{align}
\hat{\mathcal{E}}_B(\ket{\psi})\le \min_{\sigma_{aa^{\prime}b}\in\mathcal{CQ}}& d^2_B(\ket{\psi}_{ab}\bra{\psi}\otimes\ket{u}_{a^{\prime}}\bra{u},\sigma_{aa^{\prime}b})\nonumber\\
\le d^2_B(&\ket{\psi}_{ab}\bra{\psi}\otimes\ket{u}_{a^{\prime}}\bra{u},\nonumber\\
&\ket{x_1,y_1}_{ab}\bra{x_1,y_1}\otimes\ket{u}_{a^{\prime}}\bra{u})\nonumber\\
=E_B(&\ket{\psi}).
\end{align}
Combining the above two resulta, one has $\hat{\mathcal{E}}_B(\ket{\psi})=E_B(\ket{\psi})$ for any pure state $\ket{\psi}$.

On the other hand, for any mixed state with pure state decomposition $\rho_{ab}=\sum_ip_i\ket{\psi_i}_{ab}\bra{\psi_i}$, Theorem 1 tells us that
\begin{align}
\hat{\mathcal{E}}_B(\rho_{ab})\le\sum_ip_i\hat{\mathcal{E}}_B(\ket{\psi_i})=\sum_ip_iE_B(\ket{\psi_i}).
\end{align}
%Denote $E^{cr}_B(\rho_{ab})=\min_{p_i,\ket{\Psi_i}}\sum_ip_iE_B(\ket{\Psi_i})$ with the minimal over all pure state decomposition,
Taking the minimum over all pure state decompositions,
\begin{align}
\hat{\mathcal{E}}_B(\rho_{ab})\le E^{cr}_B(\rho_{ab}).
\end{align}
%As $E_B(\rho_{ab})=E^{cr}_B(\rho_{ab})$ \cite{Streltsov2010}, the theorem is proved.
\end{proof}

\subsection{Symbols and meanings}\label{symbols-and-meanings}

\begin{table}[!htbp]
\centering
\begin{tabular}{|c|c|}
\hline
\textbf{Symbol}&\textbf{Explanation}\\
\hline
$\mathcal{S}$&set of separable states\\
\hline
$\mathcal{CQ}$&set of  classical-quantum states\\
\hline
$\mathcal{CC}$&set of classical-classical states\\
\hline
$(\hat{D})~\tilde{D}$&(a)symmetric discord\\
\hline
$(\hat{D}_B)~\tilde{D}_B$&(a)symmetric  Bures distance of discord\\
\hline
$d_B$& Bures distance\\
\hline
$E_d$& geometric entanglement\\
\hline
%$d_{HS}$&Hilbert-Schmidt distance\\
%\hline
$E_B$&Bures distance of entanglement\\
\hline
$E^{cr}_B$&convex roof of Bures distance of entanglement\\
\hline
$(\hat{\mathcal{E}})~\tilde{\mathcal{E}}$ & minimal (a)symmetric DSE\\
\hline
$(\hat{\mathcal{E}}_B)~\tilde{\mathcal{E}}_B$ & minimal (a)symmetric Bures distance of DSE\\
\hline
%$\mathcal{E}^a_B(\mathcal{E}_B)$& minimal Bures distance of asymmetric (symmetric) DSE\\
%\hline
$(\hat{\mathcal{E}}_d)~\tilde{\mathcal{E}}_d$&minimal (a)symmetric DSE\\
\hline
$(\hat{\mathcal{E}}^{\prime}_d )~\tilde{\mathcal{E}}^{\prime}_d$&minimal (a)symmetric MIDSE\\
\hline
$(\check{\mathcal{E}}_d)~\breve{\mathcal{E}}_d$&minimal partial trace of (a)symmetric DSE\\
\hline
%D&Representation of $D^G_d$, $D^M_d$ or $D^L_d$\\
%\hline
$(\check{\mathcal{E}}^{\prime}_d)~\breve{\mathcal{E}}^{\prime}_d$&minimal partial trace of (a)symmetric MIDSE\\
\hline
%$D^A_{HS}$&Hilbert-Schmidt distance of asymmetric discord\\
%\hline
%$\tilde{\mathcal{E}}_{D^L_{HS}}$&Minimal $D^L_{HS}$ over cross-symmetric state extensions\\
%\hline
\end{tabular}
\caption{Symbols and their meanings. Here DSE stands for {\it discord over state extensions} and MIDSE is for {\it measurement-induced geometric discord over state extensions}.}
\end{table}


\begin{thebibliography}{99}
\bibitem{note-state-extension} We call $\rho_{aa^{\prime}b}$ an extension of quantum state $\rho_{ab}$ if it satisfies $tr_{a^{\prime}}[\rho_{aa^{\prime}b}]=\rho_{ab}$. Obviously, the purification of a quantum state is an example of state extension. By ``over state extensions" we mean that optimization runs over all extended states $\rho_{aa^{\prime}b}$ of $\rho_{ab}$.
\end{thebibliography}
\end{document}